\documentclass[twocolumn,floats,showpacs,prl]{revtex4}
\usepackage[above]{placeins}
\usepackage{amsmath}
\usepackage{graphicx}

\begin{document}

\title{Real Space Coulomb Interaction: A Pairing Glue\ for FeAs
Superconductors}
\author{Xiuqing Huang$^{1,2}$}
\email{xqhuang@nju.edu.cn}
\affiliation{$^1$Department of Physics and National Laboratory of Solid State
Microstructure, Nanjing University, Nanjing 210093, China \\
$^{2}$ Department of Telecommunications Engineering ICE, PLAUST, Nanjing
210016, China}
\date{\today}

\begin{abstract}
In this paper we present a real space pairing glue for the iron-based
layered superconductors. It is shown that two static electrons embedded
symmetrically in two adjacent Fe plaquettes of the superconductor can be
bound due to the Coulomb interaction. The pairing mechanism favors the
existence of the pseudogap in the underdoped FeAs superconductors. A
criterion is introduced to distinguish whether or not the pseudogap can open
in a material.
\end{abstract}

\maketitle

Twenty-two years after the discovery of high-$T_{c}$ cuprate superconductors
\cite{bednorz,mkwu}, although under intensive studies, physicists do not
agree on how superconductivity works in these materials \cite{anderson0}. MgB%
$_{2}$ superconductor was discovered seven years ago \cite{nagamatsu}, so
far we also do not know what causes superconducting in this material. Most
recently, the iron-based superconductors have been found by researchers in
Japan and China \cite{kamihara,xhchen} has raised the hope that the new
materials will help solve the mystery of high-$T_{c}$ superconductorson, on
the contrary, researchers now appear to be more confused than ever about the
fundamental mechanism of superconductivity. \textquotedblleft If it's really
a new mechanism, God knows where it will go.\textquotedblright\ says Philip
Anderson \cite{adrian}.

As is well known, the superconductivity is indeed a rather common phenomenon
in nature. There are now several thousand materials showing the
superconductivity. Correspondingly, physicists have developed many theories
and models on atomic level in attempts to pin down the mechanism responsible
for the observed superconducting phenomena in different materials. It seems
as if a new superconductor always calls for a new superconducting mechanism.
But this situation should not continue like that. It is time for the
condensed matter community to consider one important question: for different
superconductors, there are different superconducting mechanisms, or there is
only one unified mechanism? Personally, I think that the latter case is a
more reasonable possibility physically and naturally. The intrinsic
mechanism of the superconductivity is at base simple and determinate, it is
scientists themselves who have made the problem more complex and not
deterministic. \

Recently, we have proposed a real space spin-parallel mechanism of
superconductivity which has successfully provided coherent explanations to a
number of complicated problems in conventional and non-conventional
superconductors \cite{huang1,huang2,huang3,huang4,huang5}, for example, the
local checkerboard patterns and \textquotedblleft magic doping
fractions\textquotedblright\ in La$_{2-x}$Sr$_{x}$CuO$_{4}$ \cite{huang1},
the tetragonal vortex phase in Bi$_{2}$Sr$_{2}$CaCu$_{2}$O$_{8}$ \cite%
{huang2}, the hexagonal vortex lattice and charge carrier density in MgB$%
_{2} $ \cite{huang2}, the optimal doping phases \cite{huang5} and pressure
effects \cite{huang3} in the new iron-based superconductors, and the $%
4a\times 4a$ and $4\sqrt{2}a\times 4\sqrt{2}a$ checkerboard patterns in
hole-doped Ca$_{2-x}$Na$_{x}$CuO$_{2}$Cl$_{2}$ \cite{huang4}. Although these
results are in excellent agreement with experiments, disappointingly, they
have been totally neglected by the community.

On the other hand, despite recent attention and greater efforts to
understand the FeAs superconductors, there is no consensus on the origin of
the `superconducting glue' that binds electrons into superconducting pairs.
It should be pointed out that most of the theoretical works\ are playing
absolutely the same \textquotedblleft mathematical and numerical
games\textquotedblright\ which have been played intensively in cuprate
superconductors. Undoubtedly, many theories about electron pairing and
superconducting in the iron-based superconductors may also be on the wrong
track \cite{anderson0}. We insist that in order to have a deeper insight
into the forces responsible for Cooper pairs in the superconducting
materials, the framework of \textbf{k}-space weak-coupling BCS theory should
be abandoned and the original configuration of Cooper pairs (antiparallel
spins and opposite momenta) should be modified \cite{huang2}.

In the present paper, we aim to improve the suggested unified
superconducting theory \cite{huang1,huang2} and extend the application of
the theory to the pairing mechanism (glue) and pseudogap phase in Fe-based
superconductors.
\begin{figure*}[tbp]
\begin{center}
\resizebox{1.4\columnwidth}{!}{
\includegraphics{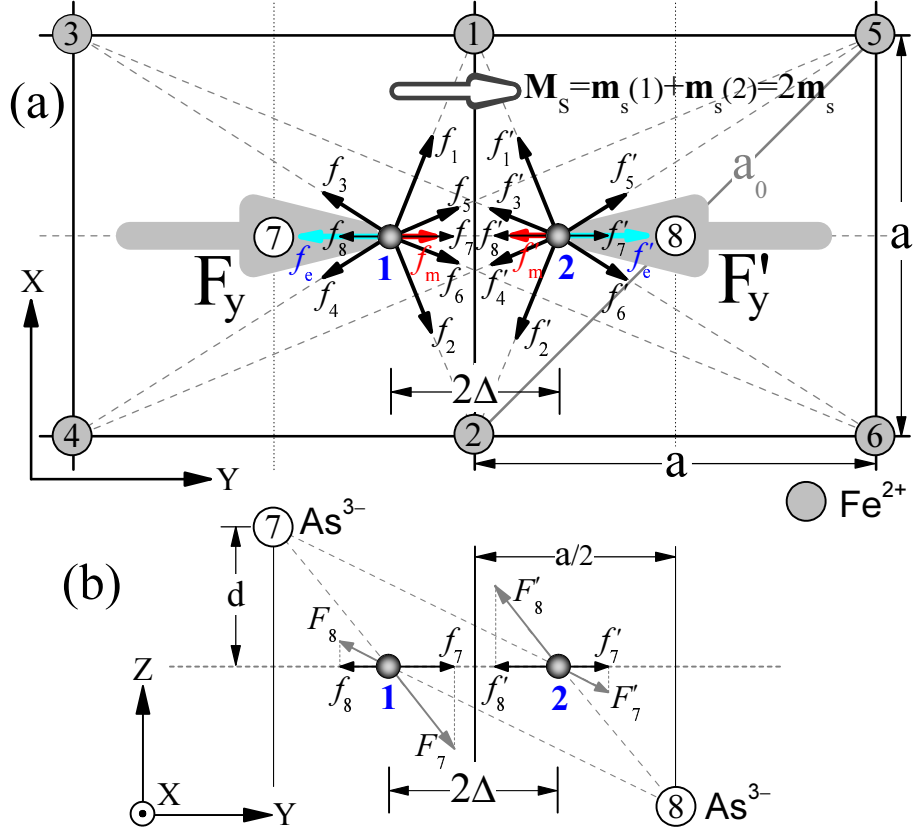}}
\end{center}
\caption{The schematic plot of the pairing glue in the FeAs superconductor.
(a) Two spin parallel electrons with a joint magnetic moment $\mathbf{M}_{s}$
is confined inside two adjacent Fe plaquettes of the superconductor. The
electron-electron ($f_{c}$ and $f_{m}$), Fe-electron ($f_{1}$, $f_{2}$, $%
f_{3}$, $f_{4}$, $f_{5}$ and $f_{6}$) and As-electron interactions ($f_{7}$
and $f_{8}$) are considered, when the net force $F_{y}=F_{y}^{^{\prime }}=0$%
, indicating a completely suppression of the Coulomb repulsion $f_{c}$ and
the opening of pseudogap, (b) picture of detailed illustration of
As-electron interactions.}
\label{fig1}
\end{figure*}

How are two negatively charged electrons bound into Cooper pairs in the
iron-based superconductors? Theoretical and numerical studies have shown
that superconductivity in these materials is associated with the FeAs layer
which can be further subdivided into a square Fe lattice with the Fe-Fe
distance $a=a_{0}/\sqrt{2}$, where $a_{0}$ is the lattice parameter. The
iron atoms are separated by arsenic atoms above and below Fe plane. As shown
in Fig. \ref{fig1}, when two static electrons embedded symmetrically into
two adjacent Fe plaquettes of the FeAs superconductor with a distance $%
2\Delta $, there is a long-range repulsive electron-electron Coulomb
interaction
\begin{equation}
f_{c}=\frac{e^{2}}{4\pi \varepsilon _{0}(2\Delta )^{2}}.  \label{fc}
\end{equation}

Obviously, the two electrons cannot be naturally paired due to this strong
repulsion. So how can repulsive Coulomb forces exerted on electrons be
eliminated so that the electrons can be in pairs? First, as shown in Fig. %
\ref{fig1}(a), for two spin parallel electrons with a joint paired-electron
magnetic moment $\mathbf{M}_{s}=\mathbf{m}_{s}(1)+\mathbf{m}_{s}(2)=2\mathbf{%
m}_{s}$ (where $\mathbf{m}_{s}$ is the monoelectron spin magnetic moment),
there is a magnetic dipolar attraction $f_{m}$ which is given by

\begin{equation}
f_{m}=\frac{3\mu _{0}\mu _{B}^{2}}{8\pi \Delta ^{4}}.  \label{fm}
\end{equation}

Because of the short-range interaction characteristics of Eq. (\ref{fm}), as
is usually the case $F_{c}\gg F_{m}^{\max }$. It is then clear that other
factors, which have the effect of weakening the long-range repulsive force $%
f_{c}$, should be taken into account. We presume that the real-space
confinement effect (electromagnetic interactions) in FeAS plane (see Fig. %
\ref{fig1}) plays a central role in suppressing the influence of the Coulomb
repulsion between electrons. For the purpose of a simplified case, we
consider the nearest-neighbor (1 and 2) Fe-electron interactions
\begin{equation}
f_{1}+f_{2}=\frac{e^{2}}{\pi \varepsilon _{0}}\frac{\Delta }{\left[
(a/2)^{2}+\Delta ^{2}\right] ^{3/2}},  \label{f12}
\end{equation}%
next-nearest-neighbor (3, 4, 5 and 6) Fe-electron interactions%
\begin{equation}
f_{3}+f_{4}=-\frac{e^{2}}{\pi \varepsilon _{0}}\frac{a-\Delta }{\left[
(a-\Delta )^{2}+a^{2}/4\right] ^{3/2}},  \label{f34}
\end{equation}%
and%
\begin{equation}
f_{5}+f_{6}=\frac{e^{2}}{\pi \varepsilon _{0}}\frac{a+\Delta }{\left[
(a+\Delta )^{2}+a^{2}/4\right] ^{3/2}}.  \label{f56}
\end{equation}%
The nearest-neighbor As-electron (7 and 8) interactions are also considered,
as shown in Fig. \ref{fig1} (b), we get
\begin{equation}
f_{7}=\frac{3e^{2}}{4\pi \varepsilon _{0}}\frac{a/2-\Delta }{\left[
(a/2-\Delta )^{2}+d^{2}\right] ^{3/2}},  \label{f7}
\end{equation}%
\begin{equation}
f_{8}=-\frac{3e^{2}}{4\pi \varepsilon _{0}}\frac{a/2+\Delta }{\left[
(a/2+\Delta )^{2}+d^{2}\right] ^{3/2}}.  \label{f8}
\end{equation}

Now we have a general formula of the total confinement force $F_{y}$ (or $%
-F_{y}^{\prime })$ applied to the electron of the pair in $y$ direction as%
\begin{eqnarray*}
F_{y} &=&-F_{y}^{\prime } \\
&=&\sum\limits_{i=1}^{8}f_{i}+f_{m}-f_{e}.
\end{eqnarray*}

\begin{figure}[tbp]
\begin{center}
\resizebox{1\columnwidth}{!}{
\includegraphics{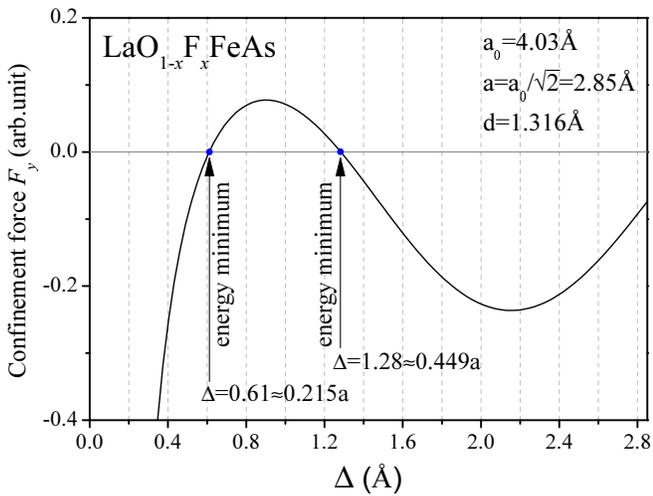}}
\end{center}
\caption{Analytical total confinement force $F_{y}$ versus $\Delta $ in LaO$%
_{1-x}$F$_{x}$FeAs superconductor. In two special positions (blue circles),
where the electrons are in the energy minimum bound states characterized by
the pseudogaps.}
\label{fig2}
\end{figure}

Physically, when $F_{y}$ (or $-F_{y}^{\prime }$) is equal to zero, it
indicates a completely suppression of the Coulomb repulsion between two
electrons. As a consequence, the electrons will be in the energy minimum
bound state. Based on the analytical expressions (\ref{fc})$-$(\ref{f8}), we
draw in Fig. \ref{fig2} the confinement force $F_{y}$ versus $\Delta $ for
LaO$_{1-x}$F$_{x}$FeAs. This figure reveals one important fact: there are
two special positions ($\Delta =0.215a$ and $0.449a$) where the localized
Cooper pair (characterized by a pseudogap) can survive in the superconductor.

\begin{figure}[tbp]
\begin{center}
\resizebox{1\columnwidth}{!}{
\includegraphics{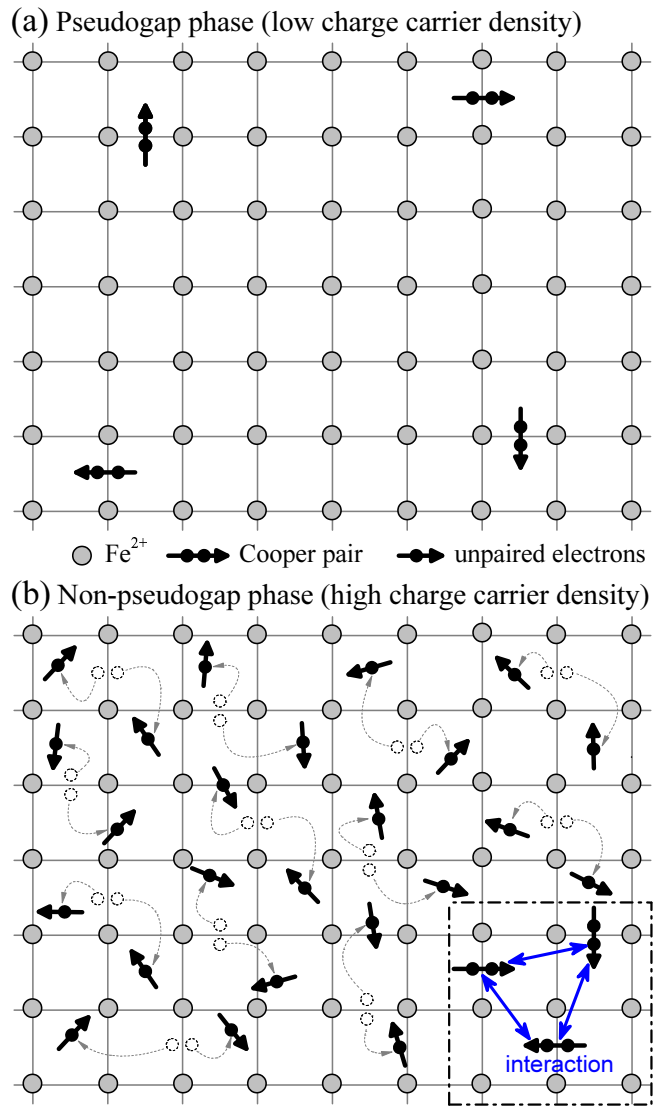}}
\end{center}
\caption{{}A schematic description of pseudogap and non-pseudogap in FeAs
superconductor. (a) At a low doping sample with a minimum pair-pair distance $%
\protect\xi >10\mathring{A}$ , the interactions among electron pairs can be
neglected, indicating the existence of the pseudogap, (b) at a high doping
concentration ($\protect\xi <10\mathring{A})$, inevitably, pseudogap phase
will be destroyed by the strong interactions among the crowded Cooper pairs.}
\label{fig3}
\end{figure}

Now we present a brief discussion of the doping dependence of the pseudogap
phenomenon. At a rather low doping level, as shown in Fig. \ref{fig3} (a),
the interactions among electron pairs can be neglected and the pairs can
maintain their integrality (pseudogap phase) at a temperature $T^{\ast }$
which is higher than the superconducting transition temperature $T_{c}$.
With an increase in doping, the effect of the competitive interactions among
pairs will emerge [see the lower right corner of Fig. \ref{fig3} (b)]. When
the doping concentration reaches certain threshold values, the localized
Cooper pairs (pseudogap phase) will be destroyed instantly due to the strong
interactions among the crowded Cooper pairs, as shown in Fig. \ref{fig3}
(b). In other words, for a given temperature, excess charge carrier
concentrations in a superconducting material is harmful for pseudogap phase.

In an earlier article \cite{huang2}, we showed that the charge carrier
density for a doped superconductor is given by
\begin{equation}
\rho _{s}=\frac{2}{ABC}=\frac{x}{abc},  \label{cd}
\end{equation}%
where $x$ is the doping level, ($a$, $b$, $c$) and ($A$, $B$, $C$) are the
constants lattice (atoms) and superlattice (electrons) constants \cite%
{huang2}, respectively. From Eq. (\ref{cd}), for a bulk superconducting
materials, we can define a parameter (the average pair-pair distance $\xi $)
as follows
\begin{equation}
\xi _{3d}=\sqrt[3]{\frac{2}{\rho _{s}}}.  \label{cd1}
\end{equation}%
For the quasi-two-dimensional layered superconductors ($c\gg a$, $b$), we
can define
\begin{equation}
\xi _{2d}=\sqrt{\frac{2}{c\rho _{s}}}=\sqrt{\frac{2ab}{x}}\approx a\sqrt{%
\frac{2}{x}}.  \label{cd2}
\end{equation}

The above parameters of Eqs. (\ref{cd1}) and (\ref{cd2}) can be used as the
criteria for the existence of pseudogap phase in the superconductors. Many
experimental results have indicated the existence of the pseudogap phases in
the underdoped cuprate and FeAs superconductors, for example, La$_{2-x}$Sr$%
_{x}$CuO$_{4}$ (LSCO, $x=0.15$) \cite{sato0} and LaO$_{1-x}$F$_{x}$FeAs
(LOFFA, $x=0.07$) \cite{sato}. However, researchers find no experimental
evidence for the pseudogap in conventional and MgB$_{2}$ superconductors.
The nature of the pseudogap phase is still highly controversial. There are
many models attempt to describe the mysterious pseudogap state. Strictly
speaking, none of the proposed models is completely satisfactory. As
discussion above, here we present a new approach based on the simple and
natural picture of the real-space confinement effect of Fig. \ref{fig1}, and
the pseudogap is associated with the local structure and the charge carrier
density in the superconductors.

According to the experimental data and Eqs. (\ref{cd}) and (\ref{cd2}), the
average pair-pair distances for La$_{2-x}$Sr$_{x}$CuO$_{4}$ ($x=0.15$) and
LaO$_{1-x}$F$_{x}$FeAs ($x=0.07$) are $\xi _{LSCO}\sim 13.88\mathring{A}$
and $\xi _{LOFFA}\sim 15.23\mathring{A}$, respectively. But MgB$_{2}$ ($%
a=b=3.086\mathring{A}$, $c=3.524\mathring{A}$ and $\rho _{s}=1.49\times
10^{22}/cm^{3}=1.49\times 10^{-2}/\mathring{A}^{3}$ \cite{huang2})
superconductor has a relatively small $\xi _{MgB}\sim 6.17\mathring{A}$,
which indicates a much strong pair-pair interaction ($\propto \xi ^{-2}$) in
the system. Here, we argue that when $\xi <10\mathring{A}$, the pair-pair
interactions are strong enough to break up the electron pairs, and
eventually closes the pseudogap in the sample. Normally, the value of the
average pair-pair distance satisfies $\xi <5\mathring{A}$ in the
conventional superconductors, thus it should not be surprising about the
non-pseudogap behavior in these materials.

In conclusion, it is found that the Coulombic interaction can play a key
role for pairing glue for the iron-based layered superconductors. The
mechanism reveals the existence of the pseudogap in a low doping FeAs sample
(underdoped), which is in satisfactory agreement with recent experiment.
Furthermore, we have introduced a criterion which can be applied to
distinguish whether or not the pseudogap can open in a material. Finally, it
should be emphasized that the suggested mechanism responsible for the
pseudogap is not specific to the iron-based family and it may also be
applicable to other superconducting and even non-superconducting materials.

\end{document}